# Optical phonons in the reflectivity spectrum of FeSi

A. Damascelli[a], K. Schulte[a], D. van der Marel[a], M. Fäth[b], A. A. Menovsky[c]

[a] Solid State Physics Laboratory, University of Groningen, Nijenborgh 4, 9747 AG Groningen, The Netherlands

[b] Department of Astronomy and Physics, University of Leiden, Nieuwsteeg 18, 2300 RA Leiden, The Netherlands

[c] Van der Waals-Zeeman Laboratory, University of Amsterdam, Valckenierstraat 67, 1018 XE Amsterdam, The Netherlands

Abstract

We measured the reflectivity of a single crystal of FeSi from the far-infrared to the visible region (50-20 000 cm$^{-1}$), varying the temperature between 4 and 300 K. The optical conductivity function was obtained via Kramers-Kronig analysis. We observed a dirty metal-like behavior at room temperature and the opening of a gap of 70 meV at low temperature. Of the five optical phonons expected from group theory analysis for this material only four have been observed in the far-infrared region.

Andrea Damascelli

Solid State Physics Laboratory

University of Groningen

Nijenborgh 4, 9747 AG Groningen

The Netherlands

Phone: ++31-50-3634922

Fax: ++31-50-3634825

Email: damascel@phys.rug.nl

FeSi is an interesting material that has been studied already many years ago for its unusual magnetic and thermal properties [1]. Nowadays this system is an object of renewed interest. During the last three years many papers, presenting both theoretical and experimental investigations, have been published [2]-[8]. FeSi is discussed as a strongly correlated system, possibly belonging to the class of materials called Kondo insulators [2]. This picture was suggested not only by magnetic susceptibility and specific heat measurements, but also by optical reflectivity data. The first infrared measurements which have been reported [7] show the opening of a gap and the transfer of spectral weight to very high frequencies with decreasing the temperature. Moreover, in contrast with what is expected from a semiconductor-like picture, the gap disappears at a temperature which is low compared to its size.

In this contribution we discuss the results of our reflectivity and optical conductivity investigations on FeSi intending to solve a lingering question [7, 8] about the nature of the bound states observed in the gap region. Therefore we present a detailed analysis of the infrared active phonon modes expected for FeSi and we compare it to the experimental data. We have measured the optical reflectivity of an FeSi single crystal of the size of approximately $2 \times 7$ mm$^2$ and thickness 1 mm, grown by the traveling floating zone method. We worked in near normal incidence configuration over the frequency range from 50 cm$^{-1}$ to 20 000 cm$^{-1}$, using two different Fourier transform spectrometers: an IFS 113v Bruker system for the far-infrared region (50 -700 cm$^{-1}$) and a DA3 Bomem system for frequencies higher than 600 cm$^{-1}$. A liquid helium flow cryostat has been used to study the temperature dependent reflectivity from 4 K to 300 K. The absolute reflectivities were obtained by calibrating the data acquired on the sample against a gold mirror from low frequency to $\sim 1\,500$ cm$^{-1}$ and an aluminum mirror for higher frequencies.

In Fig. 1 the reflectivity data and the optical conductivity obtained by Kramers-Kronig transformation [9] are displayed from 100 up to 10 000 cm$^{-1}$ in single-logarithmic scale for five different temperatures. In the insets an expanded view of the low temperature reflectivity and conductivity is shown in the frequency interval from 150 to 500 cm$^{-1}$ using a linear scale. The data are characterized by a strong temperature dependence up to 7 000 cm$^{-1}$. At room temperature the material behaves as a dirty metal: it does not show a sharp Drude peak at low frequency and the dc-conductivity $\sigma_{dc} \sim 4\,000$ $\Omega^{-1}$cm$^{-1}$ we found with dc-resistivity measurements is much smaller than the typical values for transition metals ($\sim 100\,000$ $\Omega^{-1}$cm$^{-1}$). By reducing the temperature it is possible to observe a strong depletion of the optical conductivity at low frequencies together with an onset of absorption at 570 cm$^{-1}$ (70 meV) which we identify as a semi-conducting gap.

Another evident feature is the occurrence of sharp absorption lines in the far-infrared region, and it is on this topic that we are going to concentrate our analysis in this paper. Although there has already been some discussion in previous papers on FeSi [7, 8, 11], a number of issues concerning the assignment of these peaks have remained unsolved. In Tab. 1 the resonant frequencies of modes observed with infrared and Raman spectroscopy are summarized. In the same table the results of our measurements at 4 and 300 K are also displayed. Schlesinger *et al.* [7] observed three clear peaks at 205, 332 and 466 cm$^{-1}$. They attributed all of them to phonons. Degiorgi *et al.* [8] observed peaks at 140, 200, 330, and 460 cm$^{-1}$. On the basis of group theory considerations they interpreted the 140 and 200 cm$^{-1}$ lines as phonons; for the peak at 460 cm$^{-1}$ and, possibly, for the one at 330 cm$^{-1}$ they proposed an excitonic nature, an assignment which we will change in this paper.

On the basis of group theory analysis it is possible to calculate the number and the symmetry of the lattice vibrational modes and then to determine how many optical active phonons can be expected. FeSi has a cubic structure B20 (space-group P$_{2_13}$, factor-group T(23) and site group C$_3$(4)), with four Si and four Fe atoms at equivalent positions per unit cell. This corresponds to 24 degrees of freedom, and subtracting three degrees of freedom related to the acoustical modes, we have 21 degrees of freedom associated to optical modes. Using the correlation method [10] we found for the irreducible representation of FeSi optical vibrations $\Gamma$=2A+2E+5T. As the A, E and T symmetry modes are single, doubly and triply degenerate respectively, this irreducible representation accounts for the correct number of degrees of freedom. From the lack of inversion symmetry it follows that all these modes are Raman active. But the most important conclusion is that all the five triply degenerate T symmetry modes are infrared active, and not only three as was obtained in [8]. The A and E modes are instead infrared forbidden. These results are in rather good agreement with the experimental data. In our far-infrared spectra four peaks are clearly visible (see insets of Fig. 1), with resonant frequencies at T=4 K at 207, 329, 347, 454 cm$^{-1}$ ( Tab. 1). The four resonances can all be interpreted as infrared active phonons. Moreover, while most of these peaks have already been observed [7, 8], the line at 347 cm$^{-1}$ has not been resolved before with infrared spectroscopy, possibly due to inhomogeneous broadening of the lines, as it is rather close to the 329 cm$^{-1}$ mode. There is no indication in our data for a phonon at 140 cm$^{-1}$ [8].

An additional confirmation of our group theory analysis and of the interpretation of the prominent absorption lines in the gap region as infrared active phonons is given by Raman scattering experiments done by Nyhus *et al.* [11]. They were able to measure and identify two E modes at 180 and 315 cm$^{-1}$, an A mode at 219 cm$^{-1}$ and five T modes at 193, 260, 311, 333 and 436 cm$^{-1}$ ( Tab. 1). They could resolve all the expected Raman active modes except one A mode. In particular they were able to measure all the five T modes which are expected to be both Raman and infrared active. The correspondence between Raman and infrared lines is rather good, except for the one observed by Nyhus *et al.* [11] at 260 cm$^{-1}$. A possible explanation for this difference could be a hidden symmetry of the system producing a net vanishing dipole moment for that particular oscillation mode in infrared spectroscopy.

In summary, we have investigated the optical conductivity of FeSi, noticing the opening of a gap of the order of 70 meV at low temperature. The results of a group theory analysis of the optical vibrations are reported and compared to our far-infrared data. As a final result we could assign all the sharp absorption lines observed in the gap region to phononic excitations.

# References

[1] V. Jaccarino, G. K. Wertheim, J. H. Wernick, L. R. Walker, S. Arajs, Phys. Rev. **160**, 476 (1967).

| | | | | | | |
|---|---|---|---|---|---|---|
| Schlesinger et al. [7] | - | 205 | - | 332 | - | 466 |
| Degiorgi et al. [8] | 140 | 200 | - | 330 | - | 460 |
| Nyhus et al. [11] | - | 193 | 260 | 311 | 333 | 436 |
| Our data (300K) | - | 198 | - | 318 | 338 | 445 |
| Our data (4K) | - | 207 | - | 329 | 347 | 454 |

Tab. 1. Resonant frequencies of the absorption peaks observed on FeSi in the far-infrared region with Raman (Nyhus et al.) and infrared spectroscopy (others).

Fig. 1. Reflectivity and conductivity of FeSi are shown as a function of wavenumber in single-logarithmic scale for five different temperatures. The insets show an expanded view of the 4 K data in linear scale.